# Reduction of G-factor due to Rashba Effect in Graphene

Amit Shrestha[1], Katsuhiko Higuchi[1*], Shunsuke Yoshida[2] and Masahiko Higuchi[3]

[1]*Graduate School of Advanced Sciences and Engineering, Hiroshima University, Higashi-Hiroshima 739-8527, Japan*
[2]*Graduate School of Science and Technology, Shinshu University, Matsumoto 390-8621, Japan*
[3]*Department of Physics, Faculty of Science, Shinshu University, Matsumoto 390-8621, Japan*

Graphene is a highly promising material in the field of spin electronics. Recent experiments on electron spin resonance have observed a reduction in the g-factor of graphene. In our previous paper [J. Phys. Soc. Jpn. **88**, 094707 (2019)], we demonstrated that one of sources for this reduction is the diamagnetic property of graphene. However, the diamagnetic property by itself does not fully account for the magnitude of the reduction observed in the experiments. In this paper, we focus on the Rashba effect, which is caused by the work function existing near the surface of graphene. The Rashba effect tilts the spin magnetic moment to the in-plane direction of the graphene sheet, potentially reducing the g-factor. We evaluate this reduction using a simple model system incorporating the Rashba and spin Zeeman effects. We then demonstrate that the resultant g-factor is in close agreement with that observed in the prior experiments, indicating that the Rashba effect is able to account for the remaining reduction in the g-factor of graphene.

## I. INTRODUCTION

Graphene is attracting much interest as a potential material for many applications such as electronic and spintronic devices.[1-10] The g-factor of graphene is a key quantity determining spin-related properties such as the spin relaxation time. Recent experiments on electron spin resonance (ESR)[9,10] have observed a reduction in the g-factor of graphene when subject to an external magnetic field. These experiments evaluate the g-factor by studying the effect of an external magnetic field on the gap between the lowest unoccupied state (LUS) and highest occupied state (HOS), which in graphene exists at the K point of the Brillouin zone.[11–13] Usually, this HOS-LUS gap in graphene is caused by the spin-orbit (SO) interaction, which is related to the spatially symmetric potential of the hexagonal lattice. It is shown in the previous work[14] that such an SO interaction is nearly independent of the magnetic field, and that the g-factor estimated by the differential of the HOS-LUS gap with respect to the external magnetic field[9,10] is almost equal to that of a free electron regardless of the external magnetic field.[14] However, in the experiments mentioned above,[9,10] the observed g-factor reduces by about 3.1 percent when the external magnetic field is 1 [T]. The understanding the source of this reduction in the g-factor has not only scientific value but also significance in the context of the growing interest in graphene, such as for spin electronics applications.

It is well known that graphene exhibits diamagnetism under a weak external magnetic field.[15–24] This fact has also been confirmed theoretically by means of our scheme,[14] and specifically by the magnetic-field containing relativistic tight-binding (MFRTB) method.[25,26] This method enables us to describe the electronic structure of materials immersed in a magnetic field with taking into consideration relativistic effects such as the SO interaction. In our previous paper, we demonstrated that the diamagnetism of graphene is one of the sources for the reduction in its g-factor.[14] Specifically, the estimated g-factor of graphene after accounting for the effect of diamagnetism is 1.986, when the external magnetic field is 1 [T], which is a reduction of about 0.7 percent. This, however, does not fully account for the magnitude of the reduction in the g-factor of graphene observed in previous experiments.

Another potential source for the reduction in the g-factor of grapheme is the Rashba effect.[27,28] A consideration of the Rashba effect on the graphene sheet deposited on the substrate suggests that there exists an asymmetric potential along the direction perpendicular to the graphene sheet. In the vicinity of the surface of the sheet, electrons are controlled by the work function, which spatially spreads in the region determined by electron density. Such a work function causes a magnetic field, the direction of which is in-plane along the graphene sheet, because the magnetic field is given as the cross product between the gradient of the asymmetric potential and the momentum of the electron that is in the graphene sheet.[27,28] This magnetic field is coupled to the spin magnetic moment, which forms a kind of SO interaction. As a result of this magnetic field, when an external magnetic field is applied in the direction perpendicular to the graphene sheet, the spin magnetic moment may tilt towards the in-plane direction. The extent to which the spin magnetic moment tilts depends on the strength of the Rashba effect. Therefore, in order to examine its influence on the g-factor in grapheme, in this paper we focus on the Rashba effect.

The rest of the paper is organized as follows. We first explain our scheme for calculating the g-factor of graphene, in Sec. II. Specifically, we present a procedure to calculate the reduction of the g-factor due to the Rashba effect. We then discuss an energy splitting at the K point caused by the Rashba effect (Sec. III). This energy splitting directly modifies the energy band structure of graphene, and reflects on the magnitude of the g-factor. The details of the energy band structure are described in Sec. IV. We then evaluate the reduction of the g-factor, based on a method that is appropriate for the analysis of the results of the previously mentioned experiments[9,10] on ESR (Sec. V). Finally, in the concluding section (Sec. VI), we reflect on the significance of this work and discuss issues for future work to address.

## II. CALCULATION SCHEME
### A. Calculation Procedure

In order to evaluate the extent to which the Rashba effect affects the g-factor in graphene, we perform the calculation scheme described below. In graphene, the energy gap, i.e., HOS-LUS gap, exists at the K point of the Brillouin zone. We assume that the HOS-LUS gap still exists at the K point when the Rashba effect is considered, and further when the external magnetic field is applied. In other words, we assume that the Rashba effect and external magnetic field provide only small modifications to the electronic structure of graphene. This is a reasonable assumption because the magnitude of the external magnetic field is at most about 1 [T], and the magnetic field caused by the Rashba effect is much smaller than the external magnetic field, which we demonstrate in Sec. V.

The energy bands of usual graphene possess double degeneracy at any wavenumber $\bm{k}=(k_x,k_y)$,[11–13] because it has both spatial inversion and time reversal symmetries. The energy degeneracy consists of two bands mixed by up-spin and down-spin states, and is resolved by the Rashba effect and/or the external magnetic field. This energy splitting results in the modification of the HOS-LUS gap.

(i) First the energy splitting is calculated in a model system that has a free-electron Hamiltonian with the Rashba term, spin Zeeman term and an asymmetric potential caused by the work function added. The energy splitting, whose explicit form is described in

Sec. III, is dependent on the wavenumber $k = (k_x, k_y)$. It is hereafter denoted as $\Delta_{k_x k_y}$, and its value at the K point is specifically denoted as $\Delta_K$. Note that the energy bands around the K point of the Brillouin zone in graphene can be approximated by a simple model such as an empty lattice model. We adopt the above-mentioned simple model in order to calculate the energy splitting at the K point $\Delta_K$.

(ii) Next, the obtained energy splitting $\Delta_{k_x k_y}$ is added to energy bands of the usual graphene.[13] Consequently, the original two bands, which are degenerate with each other,[5] shift by $+\Delta_{k_x k_y}$ and $-\Delta_{k_x k_y}$, respectively. In particular, the HOS and LUS are modified by $\Delta_K$. The resultant energy levels of the HOS and LUS are respectively given by

$$\varepsilon_{LUS} = \varepsilon_K + E_{so} - \Delta_K, \tag{1}$$

$$\varepsilon_{HOS} = \varepsilon_K + \Delta_K, \tag{2}$$

where $\varepsilon_K$ is the energy level of the HOS, and $E_{so}$ is the SO interaction energy caused by the spatially symmetric potential of the hexagonal lattice of the graphene sheet. The HOS-LUS gap is given by

$$\begin{aligned}\Delta\varepsilon\big|_{\text{at K}} &= \varepsilon_{LUS} - \varepsilon_{HOS} \\ &= E_{so} - 2\Delta_K.\end{aligned} \tag{3}$$

Since the reduction of the g-factor in graphene has been experimentally observed,[9,10] this reduction must be calculated by a method that can suitably enable the analysis of those ESR experimental results. To achieve this, we perform the following calculations.

(iii) From the dependence of the HOS-LUS gap on the external magnetic field, the g-factor is observed as if it were reduced.[9,10] Let the reduced g-factor be called effective g-factor. To facilitate the analysis of the experimental value observed in Refs. 9 and 10, the effective g-factor $g_{eff}$ is defined as

$$g_{eff} = \frac{1}{\mu_B} \left| \frac{d}{dB_{ext}} \Delta\varepsilon\big|_{\text{at K}} \right|, \tag{4}$$

where $\mu_B$ is the Bohr magneton and $B_{ext}$ is the external magnetic field. The details are described in Sec. V.

As mentioned above, the energy splitting caused by the Rashba and spin Zeeman effects is calculated at the K point (step (i)).  In this calculation, we use a model such that the graphene lattice is treated as an empty one.  Then, we add the thus-obtained energy splitting estimated at the K point to the HUS-LUS gap that exists at the K point of the Brillouin zone of graphene (step (ii)).  This idea of calculating the Rashba effect of graphene in our scheme is similar to that of treating the exchange-correlation (xc) effects in the local density approximation (LDA) of the density functional theory (DFT).[29–36]  In LDA, the xc effects are considered on the basis of the model of a homogeneous electron liquid, referred to as the jellium model, and the results are applied to the inhomogeneous electron system such as atoms, molecules and solids.[29]  In general terms, the idea of LDA is that xc effects are considered on the basis of a simplified model, and the results are extensively applied to more general cases.  This idea is also employed in developing the kinetic energy functional of the pair-density functional theory.[37-53]  There are other good examples such as the BCS theory, Pauli paramagnetism, Landau diamagnetism and so on.  Similarly in our present scheme, the Rashba effect is treated in a simple model system, and the results are applied to actual graphene.  Thus, the idea of this scheme is analogous to that of the LDA-DFT.

As a final note in this section, we comment briefly on the second of the two kinds of magnetic moments (the first being the spin magnetic moment) that electrons in metals primarily possess, the orbital magnetic moment.[54–57]  The orbital magnetic moment generally appears when an external magnetic field is applied to metals.  In the case of our study with graphene, the magnitude of the orbital magnetic moment for the state of the K point would be small, since the orbital of motion on the constant-energy surface of the $k$ space is point-like at the K point.[11–13]  Consequently, when solving the eigenvalue problem at the K point, we can neglect the effects of the orbital magnetic moment, and consider only the effects of the spin magnetic moment.

**B. Model of an asymmetric potential**

In this subsection, we shall explain the model for an asymmetric potential used in this paper.  Let us consider the case where graphene is placed in vacuum.  In this case, the electron at the K point is confined to the graphene sheet due to the work function that spatially spreads in the region determined by electron density.  Hereafter we refer to the potential caused by the work function as the surface potential.  When graphene is placed in vacuum, the surface potential is symmetric in the direction perpendicular to the graphene sheet.  Contributions to the energy splitting by the surface potential formed on both sides of the graphene sheet are cancelled out each other.[58]  Therefore, the energy splitting does not occur in this case.

Next, let us consider the case of the graphene sheet deposited on the substrate.  We assume that the distance between the graphene and the substrate is close enough to affect the surface potential. In this case, the surface potential is asymmetric in the direction perpendicular to the graphene sheet. This is because the surface potential formed on the substrate side is different from that on the vacuum side. The former and later surface potentials are denoted as $V_{sub}$ and $V_{vac}$, respectively. Since contributions of $V_{vac}$ and $V_{sub}$ to the energy splitting are not cancelled out each other, the energy splitting occurs in this case.

It is known that the height of $V_{sub}$ is equal to that of $V_{vac}$ minus the electron affinity of the substrate material, i.e., $V_{sub} = V_{vac} - \chi$, where $\chi$ is the electron affinity. Namely, $V_{sub}$ is smaller than $V_{vac}$ due to the electron affinity. In addition to this fact, it is shown by the experiment that the reduction of the g-factor does not largely depend on the type of substrate.[10] This experimental result seems to indicate that $V_{sub}$ is so small that it makes little contribution to the energy splitting. In this paper we assume that the contribution of $V_{sub}$ to the energy splitting is negligibly small in comparison with that of $V_{vac}$. Under this assumption, the contribution of $V_{vac}$ to the energy splitting is hardly cancelled out by that of $V_{sub}$, and therefore the energy splitting due to $V_{vac}$ almost remains. In the next section, we employ $V_{vac}$ as a model of the asymmetric potential and calculate the energy splitting $\Delta_{k_x k_y}$.

We shall give a brief comment on this model. It is well known that the large energy splitting due to the Rashba effect is observed on the Au (111) surface.[59] The experimentally-observed energy splitting is five orders of magnitude larger than the energy splitting expected from the electron gas model with the surface potential caused by the work function.[59] According to the first-principles calculation,[60] the wave function becomes asymmetric along the vertical direction to the surface due to the surface potential. The electron near the surface feels the strong potential gradient caused by the nucleus rather than that of the surface potential. This is the reason why the large energy splitting is observed on the Au (111) surface.[60] It is well known that this enhancement of the Rashba effect becomes more pronounced for heavy atoms.

The present model is based on the electron gas model with the surface potential caused by the work function. That is to say, we assume that the above-mentioned enhancement of the Rashba effect is not pronounced for the graphene sheet deposited on the substrate. This assumption may underestimate the Rashba effect. However, the Rashba effect calculated under this assumption can explain the experimentally-observed reduction of the g-factor as shown in the following sections.

## III. ENERGY SPLITTING $\Delta_{k_x k_y}$

In this paper we deal with the graphene sheet deposited on the substrate. Suppose that the direction perpendicular to the hexagonal lattice of the graphene sheet is the $z$ axis, and that the vacuum side is the positive direction.

In this section we calculate the energy splitting $\Delta_{k_x k_y}$ in a model system, the Hamiltonian of which has the Rashba term $H_{Rashba}$, spin Zeeman term $H_{Zeeman}$, and an asymmetric potential $V_{asymm}(z)$ that is caused by the work function. The Hamiltonian is

$$H = \frac{p^2}{2m} + V_{asymm}(z) + H_{Rashba} + H_{Zeeman} \tag{5}$$

where $p$ is the momentum of an electron, and $V_{asymm}(z)$ is an asymmetric potential caused by the work function, and is referred to as the surface potential in Sec. II B. As mentioned in Sec. II B, $V_{asymm}(z)$ exists in the vicinity of surface, and is supposed to be asymmetric with respect to the $z$ axis due to the existence of the substrate. The potential $V_{asymm}(z)$ also yields the Rashba effect, which is a kind of SO interaction and is expressed as the following:[27,28]

$$H_{Rashba} = \frac{g}{2} \mu_B \boldsymbol{\sigma} \cdot \boldsymbol{B}_{SO}^{asymm}(\boldsymbol{r}), \tag{6}$$

with

$$\boldsymbol{B}_{SO}^{asymm}(\boldsymbol{r}) = \frac{2}{g\mu_B} \frac{\hbar}{4m^2 c^2} \nabla V_{asymm}(z) \times \boldsymbol{p}, \tag{7}$$

where $\boldsymbol{\sigma}$ is the Pauli matrix, $g$ is the conventional g-factor given by $g = 2.0023$, and $\mu_B$, $\hbar$, $m$ and $c$ are the Bohr magneton, reduced Planck constant ($\hbar = h/(2\pi)$), rest mass of an electron and speed of light in vacuum, respectively. Equation (7) is sometimes called the Rashba magnetic field. If the Rashba parameter is defined as

$$\boldsymbol{\alpha}_{asymm}(\boldsymbol{r}) = \frac{\hbar}{4m^2 c^2} \nabla V_{asymm}(z), \tag{8}$$

then the Rashba term Eq. (6) is rewritten as

$$H_{Rashba} = \boldsymbol{\sigma} \cdot \{\boldsymbol{\alpha}_{asymm}(\boldsymbol{r}) \times \boldsymbol{p}\}. \tag{9}$$

Suppose that the external magnetic field $\boldsymbol{B}_{ext}$ is applied in the $z$ direction. i.e., $\boldsymbol{B}_{ext} = (0, 0, B_{ext})$. The spin Zeeman effect with respect to the external magnetic field is given by

$$H_{Zeeman} = \frac{g}{2} \mu_B \sigma_z B_{ext}. \tag{10}$$

In order to diagonalize the Hamiltonian Eq. (5), we choose the eigenfunctions of a subsystem of Eq. (5) as the basis functions. Specifically, as a subsystem, let us consider the Hamiltonian excluding the third and fourth terms from Eq. (5), and denote it as $H_0$. The subsystem $H_0$ is written as

$$H_0 = -\frac{\hbar^2}{2m}\left(\frac{\partial^2}{\partial x^2} + \frac{\partial^2}{\partial y^2} + \frac{\partial^2}{\partial z^2}\right) + V_{asymm}(z). \tag{11}$$

If the eigenfunction and eigenvalue for $H_0$ are denoted as $\varphi_{k_x k_y n}(\boldsymbol{r})$ and $\varepsilon_{k_x k_y n}$, respectively, the subsystem obeys the following equations:

$$H_0 \varphi_{k_x k_y n}(\boldsymbol{r}) = \varepsilon_{k_x k_y n} \varphi_{k_x k_y n}(\boldsymbol{r}), \tag{12}$$

with

$$\varphi_{k_x k_y n}(\boldsymbol{r}) = \frac{1}{\sqrt{L^2}} e^{i(k_x x + k_y y)} u_n(z), \tag{13}$$

and

$$\varepsilon_{k_x k_y n} = \frac{\hbar^2}{2m}\left(k_x^2 + k_y^2\right) + \xi_n, \tag{14}$$

where $u_n(z)$ and $\xi_n$ satisfy the equation

$$\left\{-\frac{\hbar^2}{2m}\frac{\partial^2}{\partial z^2}+V_{asymm}(z)\right\}u_n(z)=\xi_n u_n(z). \tag{15}$$

When diagonalizing the Hamiltonian Eq. (5), we utilize the set of $\varphi_{k_x k_y n}(\mathbf{r})$ as the basis set. Here we employ an approximation such that the Rashba parameter Eq. (8) is spatially constant. Specifically, $\boldsymbol{\alpha}_{asymm}(\mathbf{r})$ is approximated into $\tilde{\alpha}_z \mathbf{e}_z$, where $\mathbf{e}_z$ is a unit vector of the $z$ axis. This approximation implies that the electric field in the $z$ direction, i.e., $\nabla V_{asymm}(z)$, is assumed to be constant.[61] The approximation corresponds to the assumption that $V_{asymm}(z)$ is linearly proportional to $z$ in the vicinity of the surface, which is reasonable.[61] Further, this approximation does not disturb the original symmetry of the system given by Eq. (5) in which the spatial inversion symmetry does not exist. With this approximation, the matrix elements of Eq. (5) are given by

$$\begin{aligned}H_{k_x k_y n\sigma, k'_x k'_y n'\sigma'}=&\left\{\frac{\hbar^2}{2m}\left(k_x^2+k_y^2\right)+\xi_n\right\}\delta_{k_x k'_x}\delta_{k_y k'_y}\delta_{nn'}\delta_{\sigma\sigma'}\\ &+\left\{-\tilde{\alpha}_z \hbar k_y \langle\chi_\sigma|\sigma_x|\chi_{\sigma'}\rangle+\tilde{\alpha}_z \hbar k_x \langle\chi_\sigma|\sigma_y|\chi_{\sigma'}\rangle\right\}\delta_{k_x k'_x}\delta_{k_y k'_y}\delta_{nn'}\\ &+\frac{g}{2}\mu_B B_{ext}\langle\chi_\sigma|\sigma_z|\chi_{\sigma'}\rangle\delta_{k_x k'_x}\delta_{k_y k'_y}\delta_{nn'}.\end{aligned} \tag{16}$$

The diagonalization can be performed in each block matrix with the same $(k_x, k_y)$ and $n$. The block matrix is a $2\times 2$ matrix, and is given by

$$h(k_x,k_y,n)=\begin{pmatrix}\frac{\hbar^2}{2m}\left(k_x^2+k_y^2\right)+\xi_n+\frac{g}{2}\mu_B B_{ext} & -\tilde{\alpha}_z \hbar\left(k_y+ik_x\right)\\ -\tilde{\alpha}_z \hbar\left(k_y-ik_x\right) & \frac{\hbar^2}{2m}\left(k_x^2+k_y^2\right)+\xi_n-\frac{g}{2}\mu_B B_{ext}\end{pmatrix}. \tag{17}$$

The eigenvalue of the block matrix $h(k_x,k_y,n)$ is easily obtained. We have

$$E_\pm(k_x,k_y,n)=\varepsilon_{k_x k_y n}\pm\Delta_{k_x k_y}, \tag{18}$$

where

$$\Delta_{k_x k_y} = \sqrt{\delta(B_{ext})^2 + (\tilde{\alpha}_z \hbar)^2 (k_x^2 + k_y^2)}, \qquad (19)$$

and

$$\delta(B_{ext}) = \frac{g}{2} \mu_B B_{ext}. \qquad (20)$$

The energy splitting $\Delta_{k_x k_y}$ consists of both an energy splitting due to the spin Zeeman effect, i.e., $\delta(B_{ext})$, and energy splitting due to the Rashba effect, i.e., $(\tilde{\alpha}_z \hbar)^2 (k_x^2 + k_y^2)$. If the magnitude of the former is much larger than that of the latter, the energy splitting $\Delta_{k_x k_y}$ is approximated into

$$\Delta_{k_x k_y} \approx |\delta(B_{ext})| + \frac{(\tilde{\alpha}_z \hbar)^2 (k_x^2 + k_y^2)}{2|\delta(B_{ext})|}. \qquad (21)$$

As will be shown in Sec. V, the magnitude of the Rashba magnetic field, i.e., magnitude of Eq. (7), is about one-fourth of that of the external magnetic field if the latter is 1 [T]. Therefore, the assumption that Eq. (21) holds is appropriate for the case of graphene.

## IV. ENERGY BAND STRUCTURE

In the Hamiltonian Eq. (5), the spatial inversion symmetry is lost due to an asymmetric potential and/or Rashba term. In addition, the time reversal symmetry is lost due to the spin Zeeman term. However, the rotational symmetry with respect to any angle around the $z$ axis is preserved in Eq. (5). As a result, the energy splitting $\Delta_{k_x k_y}$ is invariant under a transformation that does not change the magnitude of $k = (k_x, k_y)$. In other words, the energy splitting $\Delta_{k_x k_y}$ is symmetric with respect to the origin of $k$. This is easily confirmed in Eq. (21).

In our present scheme, as mentioned in Sec. II, the energy splitting $\pm\Delta_{k_x k_y}$ are simply added to the original energy bands of usual graphene.[13] As a result, the original degenerate bands, which correspond to two bands mixed by up-spin and down-spin bands, shift by $+\Delta_{k_x k_y}$ and $-\Delta_{k_x k_y}$, respectively. The schematic diagram of energy bands around the K point is illustrated in Fig. 1. Figure 1 reveals that the HOS, LUS and HOS-LUS gap are given by Eqs. (1), (2) and (3), respectively. Substituting Eq. (21) into Eq. (3), the HOS-LUS gap of graphene is given by

$$\Delta\varepsilon\Big|_{at\ K} = E_{so} - \left\{ 2|\delta(B_{ext})| + \frac{(\tilde{\alpha}_z \hbar)^2 (k_x^2 + k_y^2)}{|\delta(B_{ext})|} \right\}\Bigg|_{at\ K}. \qquad (22)$$

The HOS-LUS gap is modified by the second term of the RHS of Eq. (22). The second term directly affects the reduction of the g-factor, since it is dependent on the external magnetic field.

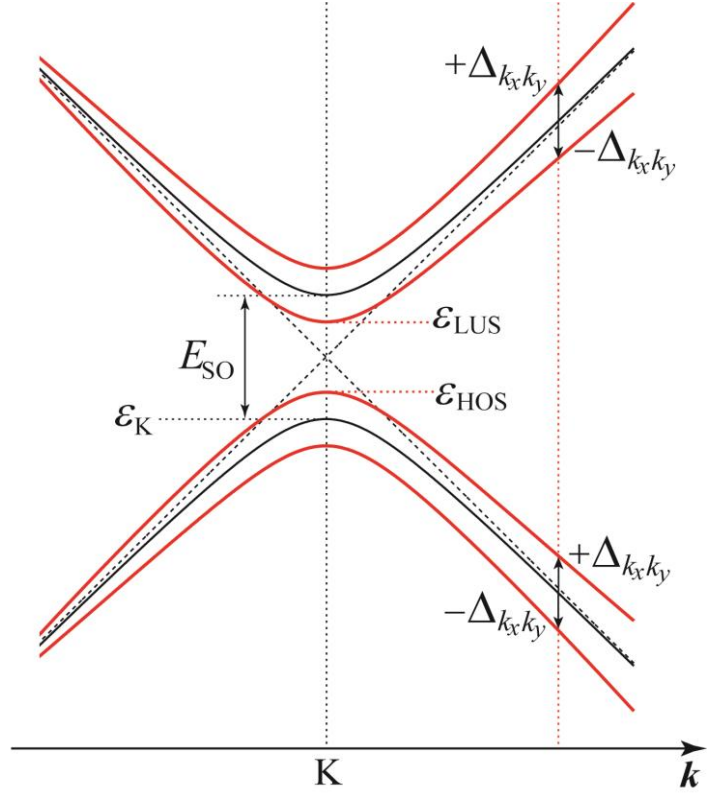

Fig. 1: Schematic diagram of energy bands around the K point. Solid lines indicate degenerated bands mixed by up-spin and down-spin states. Red lines indicate the bands shifted by the Rashba effect and external magnetic field. Dashed lines indicate the energy bands in the case neglecting the SO interaction.

## V. EFFECTIVE G FACTOR
### A. Estimation of the effective g-factor

In the previous experiments on ESR,[9,10] the effective g-factor is derived using the dependence of the HOS-LUS gap on the external magnetic field. Specifically, in these experiments, $\Delta\varepsilon|_{at\ K}$ is regarded as the spin Zeeman energy, and the proportional coefficient to the external magnetic field is derived by differentiating $\Delta\varepsilon|_{at\ K}$ with respect to the external magnetic field. This method of derivation is different from the one that has been previously used in the literature, in which the SO interaction energy $E_{so}$ explicitly changes the value of the effective g-factor,[62] while in this method $E_{so}$ does not affect the effective g-factor.[9,10] It is therefore possible to exclude the effect of $E_{so}$ from the effective g-factor in this derivation. Thus, the effective g-factor is defined as Eq. (4). Substituting Eq. (22) into Eq. (4), we get

$$g_{eff} = g \left\{ 1 - 2\left(\frac{\tilde{\alpha}_z \hbar}{g\mu_B B_{ext}}\right)^2 \left(k_x^2 + k_y^2\right)\bigg|_{at\,K} \right\}. \tag{23}$$

The expression Eq. (23) can also be derived in a more intuitive manner. In order to derive the tilting angle of the spin magnetic moment from the $z$ axis to the $x-y$ plane, it is sufficient to calculate the ratio of the Rashba magnetic field $\tilde{B}_{so}^{asymm}$ to the external magnetic field $B_{ext}$. Here, $\tilde{B}_{so}^{asymm}$ is defined as an approximation of $\left|\boldsymbol{B}_{SO}^{asymm}(\boldsymbol{r})\right|$ given by Eq. (7) such that $\boldsymbol{p}$ and $\boldsymbol{\alpha}_{asymm}(\boldsymbol{r})$ are approximated into $\hbar\boldsymbol{k} = \hbar(k_x, k_y)$ and $\tilde{\alpha}_z \boldsymbol{e}_z$, respectively. That is, we have

$$\tilde{B}_{so}^{asymm} = \frac{2}{g\mu_B}|\tilde{\alpha}_z|\hbar\sqrt{k_x^2 + k_y^2}. \tag{24}$$

The ratio of $\tilde{B}_{so}^{asymm}$ to $B_{ext}$ gives the tilting angle of the spin magnetic moment from the $z$ axis to the $x-y$ plane. The schematic view of the Rashba magnetic field, external magnetic field and spin magnetic moment is given in Fig. 2. If the angle is denoted as $\theta$, we have $\tan\theta = \tilde{B}_{so}^{asymm}/B_{ext}$. Using Eq. (24), the tilting angle is rewritten as

$$\tan\theta = \frac{2|\tilde{\alpha}_z|\hbar\sqrt{k_x^2 + k_y^2}}{g\mu_B B_{ext}} \tag{25}$$

The spin magnetic moment is substantially reduced by the factor $\cos\theta$. If the tilting angle is small, such that the Rashba effect is smaller than the spin Zeeman effect, then $\cos\theta$ is given by

$$\cos\theta \approx 1 - 2\left(\frac{\tilde{\alpha}_z \hbar}{g\mu_B B_{ext}}\right)^2 \left(k_x^2 + k_y^2\right) \tag{26}$$

We thus obtain the tilting angle of the spin magnetic moment. Comparing Eq. (26) to Eq. (23), we can see that the reduction of the g-factor is also explained using the ratio of the Rashba magnetic field to the external magnetic field. Alternatively, we can also confirm Eq. (26) by calculating the expectation value of the spin magnetic moment with respect to the eigenfunctions of Eq. (17). This shows that the spin magnetic moment of the HOS tilts by

$\theta$ at the K point, while that of the LUS tilts by $\pi-\theta$ at the K point.

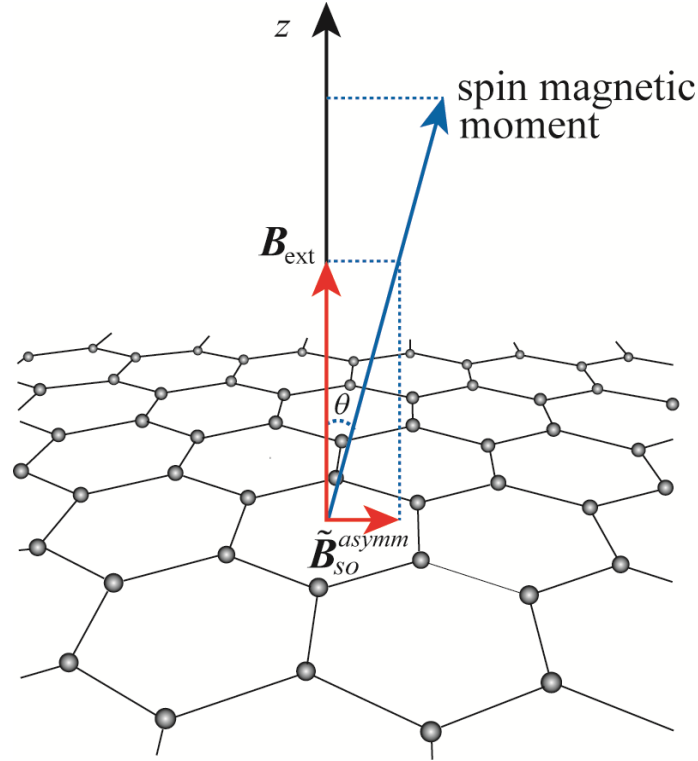

Fig. 2: Schematic view of the Rashba magnetic field, external magnetic field and spin magnetic moment. The tilting angle $\theta$ of the spin magnetic moment is given by Eq. (26).

In order to evaluate the effective g-factor quantitatively, the value of the Rashba parameter $\tilde{\alpha}_z$ is needed, as shown in Eq. (23). We calculate $\tilde{\alpha}_z$ in the following three steps, and using this we derive the effective g-factor.

*Step 1*:

In order to calculate the Rashba parameter $\tilde{\alpha}_z$, we need the electric field $\nabla V_{asymm}(z)$, as can be seen in Eq. (8). An asymmetric potential $V_{asymm}(z)$ is caused by the work function. The work function generally consists of the surface term and the bulk term.[61] The surface term originates from the electrical double layer existing near the surface. Electrons seep into the vacuum side from the surface due to the tunnel effect, due to which the electrical double layer forms in the vicinity of the surface. The electrons cannot easily escape from the surface, which is an origin of the work function.[61] The bulk term corresponds to the difference between the total energy of isolated atoms and that of the jellium model of electron liquid. This term therefore coincides with the cohesive energy times the factor -1.[61]

In this paper, we use the experimental value $W = 4.3\,[\text{eV}]$, obtained from,[63] for the work function.

*Step 2*:

The work function spatially spreads over some region near the surface. Correspondingly, electron density may decrease in that region because the work function acts as a potential barrier for the electrons. Therefore, the decay length of the electron density can be regarded as the spread length of the work function. In this step, we evaluate the decay length of the electron density in the vicinity of the surface of graphene using the jellium model of the electron liquid.[61] Let this decay length be referred to as $d$.

In the paper by Lang-Kohn,[61] the spatial profile of the electron density is derived for various $r_s$ parameters. In Appendix, we calculate the $r_s$ parameter of the graphene sheet, using the effective thickness of the graphene sheet. This results in a value of $r_s = 1.93548$ for the graphene sheet. Accordingly, we borrow a result from prior work for the spatial profile of electron density of $r_s = 2$.[61] In such a profile,[61] assume that a length decreasing the electron density from $0.95\bar{n}_p$ to $0.05\bar{n}_p$ is regarded as the decay length of the electron density, where $\bar{n}_p$ is the uniform positive charge density of the jellium model. The resulting $d$ is $1.7604 \times 10^{-10}\,[m]$. As mentioned above, this value of $d$ corresponds to the spread length of the work function.

The electric field caused by the work function is supposed to be constant in our present scheme, as mentioned in Sec. III. Using the work function $W$ and its spread length $d$, the asymmetric potential is written as

$$V_{asymm}(z) = \frac{W}{d} z . \tag{27}$$

Using Eqs. (8) and (27), the Rashba parameter $\tilde{\alpha}_z$ is given by

$$|\tilde{\alpha}_z| = \frac{\hbar}{4m^2 c^2} \frac{W}{d}, \tag{28}$$

where $W = 4.3\,[\text{eV}]$ and $d = 1.7604 \times 10^{-10}\,[m]$ as shown in *Step 1* and *Step 2*, respectively.

As a reference, the corresponding Rashba magnetic field $\tilde{B}_{so}^{asymm}$ is calculated to be 0.268 [T], which is much smaller than the magnitude of the external magnetic field $B_{ext} = 1\,[T]$. This implies that the premise of Eq. (21) is correct. The relation between the Rashba magnetic field, external magnetic field and spin magnetic moment is illustrated in Fig. 2.

Let us compare the magnitude of the Rashba effect with previous estimations.[13,64,65]

The magnitude of the Rashba effect is usually evaluated by the energy splitting of the Dirac cones in the case where the electric field of 1 [V/nm] is applied perpendicularly to the graphene sheet. This energy splitting is usually denoted as $2\lambda_R$. Estimated previous and present values of $2\lambda_R$ are summarized in TABLE I. The value of $2\lambda_R$ for graphene was first estimated by Kane and Mele as 0.516 [µeV nm/V].[64] Min *et al.* estimated it as 133.2 [µeV nm/V] by the tight-binding approximation method.[65] The estimation of $2\lambda_R$ through the first-principles calculation is done by Gmitra *et al.*, and they estimated it as 9.9 [µeV nm/V].[13] In order to compare these estimations with our estimation, we consider the energy splitting due to the Rashba effect in the case of $B_{ext} = 0$. The energy splitting for the case of $B_{ext} = 0$ is calculated by using Eqs. (19) and (28). The estimated value of $2\lambda_R$ is 1.27 [µeV nm/V]. This value is about 2.5 times larger than Kane and Mele's estimation, and is about 1/8 and 1/105 of the estimations by Min *et al.* and Gmitra *et al.*, respectively. Although previous numerical estimations for $2\lambda_R$ are rather controversial, our value is about in the middle of them. As described in the next step (step 3), the present estimation of energy splitting can successfully explain the experimental results of g-factor.

TABLE I. Estimated values of $2\lambda_R$. Here $2\lambda_R$ denotes the energy splitting of the Dirac cones due to the Rashba effect in the case where the electric field of 1 [V/nm] is applied perpendicularly to the graphene sheet.

|  | $2\lambda_R$ [µeV nm/V] |
| --- | --- |
| Kane and Mele[64] | 0.516 |
| Min *et al.*[65] | 133.2 |
| Gmitra *et al.*[13] | 9.9 |
| Present work | 1.27 |

*Step 3*:

Finally, in order to evaluate the effective g-factor, we need the magnitude of $k$ at the K point. The K point and its magnitude are given by $k|_{at\,K} = (2\pi/a)(1/3, 1/\sqrt{3})$, and $k|_{at\,K} = 1.70276 \times 10^{10}$ [$m^{-1}$], respectively. Substituting the results obtained in these three steps into Eq. (23), with an external magnetic field of $1[T]$, we get $g_{eff} = 1.931$ (see TABLE II). This is the effective g-factor predicted by our scheme at $B_{ext} = 1[T]$.

This reveals that the reduction of g-factor is actually caused by the Rashba effect. The g-factor is reduced by about 3.6 percent in our scheme, while its experimentally observed value is a reduction by about 3.1 percent[9] or 2.5 percent[10] when the external magnetic field is

1 [T] (TABLE II). This establishes the fact that the Rashba effect is a primary cause for the reduction of the g-factor in graphene. As already explained in Sec. I, the diamagnetism of graphene is another source for the reduction of its g-factor, by about 0.7 percent.[14] Additionally accounting for this factor suggests that our scheme slightly overestimates the influence of the Rashba effect on the g-factor of graphene. In the subsequent section, we comment on the possibility of other influencing factors that might be causing this overestimation in our scheme.

TABLE II. Effective g-factor at $B_{ext} = 1[T]$.

|  | effective g-factor |
| --- | --- |
| Mani et al.[9] (experiment) | $1.94 \pm 0.024$ |
| Lyon et al.[10] (experiment) | $1.952 \pm 0.002$ |
| Present work | 1.931 |

At the end of this section, we give a brief comment on the case where the external magnetic field is applied parallel to the graphene sheet. The effective magnetic field $B_{SO}^{asymm}(r)$ caused by the Rashba effect is parallel to the graphene sheet as shown in Fig. 2, irrespective of the direction of the external magnetic field. Namely, $B_{SO}^{asymm}(r)$ is parallel to the external magnetic field in this case. Although the HOS-LUS gap changes with the external magnetic field, the differential of the HOS-LUS gap with respect to the external magnetic field would not change. Therefore, the g-factor is expected to be close to that of a free electron, which is consistent with the earlier experimental works of graphite.[66,67]

**B. Gate voltage dependence of the effective g-factor**

It is shown by the experiment on ESR[10] that the effective g-factor does not depend on the gate voltage, where the externally-applied electric-filed ($E_z^{ext}$) perpendicular to the graphene sheet is controlled by the gate voltage. We consider the inherently-existing electric filed ($E_z^{in}$) that comes from the work function as mentioned in preceding sections. When we consider the gate voltage dependence of the energy splitting due to the Rashba effect, we should take both $E_z^{ext}$ and $E_z^{in}$ into consideration. The magnitude of $E_z^{in}$ is estimated as $E_z^{in} = 24.4$[V/nm] by using Eq. (27), $W = 4.3$ [eV] and $d = 1.7604 \times 10^{-10}$ [m]. On the other hand, $E_z^{ext} \sim 50 \text{V} / 300 \text{nm}$ is often used in estimating the Rashba effect caused by the externally-applied electric-filed.[13,64,65] This value is more than two orders of magnitude smaller than $E_z^{in}$. It is expected that $E_z^{ext}$ is much smaller than $E_z^{in}$ even if the gate voltage changes from -30 [V] to 20 [V] like in the experiment on ESR.[10] Therefore, the Rashba effect in the graphene sheet deposited on the substrate is mainly caused by $E_z^{in}$. It is expected from the present scheme that the energy splitting as well as the effective g-factor is almost independent of the gate voltage. Thus, the present scheme can describe the experimental fact that the effective g-factor is almost independent of the gate voltage.

## VI. CONCLUDING REMARKS

In this paper, we set out to explain the reduction of the g-factor of graphene[9,10] by focussing on the Rashba effect, which is caused by the work function existing near the surface. We derive the HOS-LUS gap on the basis of a model system, and then calculate the effective g-factor.

The scheme described in this paper satisfactorily explains that a primary cause for the reduction in the g-factor in graphene is the Rashba effect. The Rashba effect tilts the spin magnetic moment towards the in-plane direction of the graphene sheet, such that energy splitting by the external magnetic field substantially decreases. This leads to a reduction in the observed g-factor in the graphene sheet.

As mentioned in Sec. II B, the present model may possibly underestimate the Rashba effect. If the Rashba effect was enhanced in the graphene sheet due to the potential gradient of the nucleus, then the g-factor would become much smaller than that of the present work. Thus, even though the Rashba effect is small for the graphene sheet on the substrate, it can explain the experimental results of the g-factor. It seems that the measurement of the g-factor by ESR can be an effective method for estimating the small Rashba effect.

However, there also appears to be a quantitative error in the effective g-factor obtained from our scheme. Specifically, the scheme overestimates the reduction of the g-factor by 0.5 percent in comparison with the experimental value.[9,10] This overestimation of the Rashba effect may be improved by considering the effect of the weakened surface potential on the substrate side ($V_{sub}$) that is assumed to be negligibly small (Sec. II B). It will be a future work to evaluate the effect of $V_{sub}$. Besides the Rashba effect, there are other effects that might potentially modify the effective g-factor as follows:

(i) The orbital magnetic moment generally appears in metals when the external magnetic field is applied to the system. As mentioned Sec. II, the magnitude of the orbital magnetic moment for the state of the K point is likely to be small in graphene, due to which it was ignored in our scheme. Future work can examine the actual magnitude of effect of the orbital magnetic moment on the g-factor to confirm whether it is significant or not. The effect of the orbital magnetic moment can be treated theoretically by using the MFRTB method.[14,25,26] In addition, the relativistic effects such as the SO interaction are taken into consideration by using the MFRTB method.[14,25,26] Our subsequent work will consider extending the MFRTB method[14,25,26] to incorporate the Rashba effect, in order to apply it on graphene.

(ii) In metals, besides the external magnetic field, there exists an internal magnetic field that is generated by the motions of electrons in the system. Diamagnetism is one such kind of internal magnetic field. A feasible approach to determining the internal magnetic field might be to calculate the diamagnetism using the dependence of the total energy on the external magnetic field. While this approach is feasible with the MFRTB method, and we have utilized it to estimate diamagnetism of graphene in our previous work,[14] the MFRTB method does not incorporate the Rashba effect. Therefore, in this paper, as mentioned in Sec. V, we have only referred to our previous result on the diamagnetism of graphene.[14] A next challenge to achieve this could be to extend the MFRTB method to incorporate the Rashba effect, and then re-evaluate the diamagnetism of graphene with the new method.

We have discussed the above ideas as avenues for future work to explore, in order to achieve a finer numerical agreement with the experimentally recorded value of the g-factor of graphene, and to achieve further enhancement of the calculation scheme. Nevertheless, the work in this paper has demonstrated that the Rashba effect is a major source for the reduction in the g-factor of graphene, accounting for most of this reduction. Our findings establish that, from a quantitative viewpoint, the Rashba effect plays an indispensable role in the reduction of the g-factor of graphene.

The g-factor of graphene has been a topic of great interest in upcoming application areas such as spin electronics because it determines the spin relaxation time.[9,10] Therefore, our findings provide a useful reference in developing spintronics devices. Further, the Rashba effect may cause the reduction of the g-factor also in other atomic layer materials. For example, the bilayer graphene is attractive for application areas due to the opening of the energy band gap.[68] According to the experiment,[69] the work function of the bilayer graphene is about 4.4-4.5 [eV], which is slightly larger than that of the monolayer graphene. Since the difference is small, the Rashba effect caused by the surface potential in the bilayer graphene might not be much different from that in the monolayer graphene. Our achievements of this work can generally be a useful reference when investigating the g-factor of atomic layer materials.

**APPENDIX:** Effective thickness and $r_s$ parameter of graphene

In this appendix, we present the effective thickness of graphene by considering the magnetization of graphene. Further, we use it to calculate the $r_s$ parameter of graphene, i.e., the electron density of graphene.

First, we derive the effective thickness of graphene, which is referred to as $l$. In our previous work,[14] the sheet magnetization, which corresponds to the magnetization per unit area, is defined as

$$M_{sheet} = -\frac{dE_{tot}}{dB_{ext}}, \tag{A1}$$

where $E_{tot}$ is the total energy of graphene per unit area. If the magnetic dipole moment per one carbon atom in graphene is denoted as $m_{carbon}$, it is given by

$$m_{carbon} = S_{carbon}\mu_0 M_{sheet}, \tag{A2}$$

where $S_{carbon}$ is the area occupied by one carbon atom in graphene. That is given by

$$S_{carbon} = \frac{\sqrt{3}}{4}a^2, \tag{A3}$$

where $a$ is the lattice constant of graphene.

On the other hand, if we assume that $m_{carbon}$ is caused by the electric current $I_{carbon}$ flowing along the edge of the triangular cell which contains one carbon atom, $m_{carbon}$ is written as

$$m_{carbon} = \mu_0 I_{carbon} S_{carbon}. \tag{A4}$$

Comparing Eq. (A4) to Eq. (A2), we get the relation

$$M_{sheet} = I_{carbon}. \tag{A5}$$

From the Biot-Savar law, the magnetic field that is formed by $I_{carbon}$ at the position of the carbon atom is given by

$$B_{ind} = \frac{9\mu_0}{2\pi}\frac{M_{sheet}}{a}, \tag{A6}$$

where Eq. (A5) is used. We assume that the magnetic field in the effective region $S_{carbon}l$ is represented by this magnetic field $B_{ind}$. Considering that $B_{ind}$ originates from $M_{sheet}$, the relation $B_{ind} = \mu_0 M_{sheet}/l$ holds. Using this relation and Eq. (A6), the effective thickness of graphene is written as

$$l = \frac{2\pi a}{9}. \tag{A7}$$

Using the effective thickness of graphene, we can calculate the electron density. In graphene, one conduction electron (sometimes referred to as $\pi$-electron) is supplied per carbon atom. The volume occupied by one carbon atom is given by $\Omega_{carbon} = S_{carbon}l$. The electron density of graphene is given by $n = 1/\Omega_{carbon}$. Substituting Eqs. (A3) and (A7) into this equation, we get

$$n = \frac{6\sqrt{3}}{\pi a^3}. \tag{A8}$$

The $r_s$ parameter is defined as

$$r_s = \frac{1}{a_B}\left(\frac{3}{4\pi n}\right)^{\frac{1}{3}}, \tag{A9}$$

where $a_B$ is the Bohr radius. Substituting Eq. (A8) into Eq. (A9), and using the lattice constant of graphene: $a = 2.46 \times 10^{-10}$ [m], we finally get the $r_s$ parameter of graphene: $r_s = 1.93548$.

**ACKNOWLEGEMENTS**

This work was partially supported by Grant-in-Aid for Scientific Research (No. 18K03510 and No. 18K03461) of the Japan Society for the Promotion of Science.

**DATA AVAILABILITY**

The data that support the findings of this study are available from the corresponding author upon reasonable request.